\documentclass[twocolumn]{aastex62}

\usepackage{comment}
\usepackage{ulem}
\usepackage{tikz}
\usepackage{color}
\usepackage{graphicx}
\usepackage{amsmath}
\usepackage{amsfonts}
\usepackage{outlines}
\usepackage{enumitem}

\newcommand{\beq}{\begin{equation}}
\newcommand{\eeq}{\end{equation}}
\newcommand{\bea}{\begin{eqnarray}}
\newcommand{\eea}{\end{eqnarray}}

\newcommand{\gv}[1]{\ensuremath{\mbox{\boldmath$ #1 $}}} 
\newcommand{\trm}[1]{\textrm{#1}}

\newcommand{\non}{\nonumber \\}

\newcommand{\muHz}{\trm{ $\mu$Hz}}

\DeclareGraphicsExtensions{.png,.pdf}

\begin{document}

\shortauthors{WEINBERG \& ARRAS}
\shorttitle{Nonlinear Mixed Modes}

\title{\bf \large Nonlinear Mixed Modes in Red Giants}

\author{Nevin~N.~Weinberg}
\affiliation{Department of Physics, and Kavli Institute for Astrophysics and Space Research, \\Massachusetts Institute of Technology, Cambridge, MA 02139, USA}

\author{Phil~Arras}
\affiliation{Department of Astronomy, University of Virginia, P.O. Box 
400325, Charlottesville, VA 22904-4325, USA}


\begin{abstract}
Turbulent motions in the convective envelope of red giants excite a rich spectrum of solar-like oscillation modes.  Observations by \textit{CoRoT} and \textit{Kepler} have shown that the mode amplitudes increase dramatically as the stars ascend the red giant branch, i.e., as the frequency of maximum power, $\nu_\mathrm{max}$, decreases.  Most studies nonetheless assume that the modes are well described by the linearized fluid equations.  We investigate to what extent the linear approximation is justified as a function of stellar mass $M$ and $\nu_\mathrm{max}$,  focusing on dipole mixed modes with frequency near $\nu_\mathrm{max}$.  A useful measure of a mode's nonlinearity is the product of its radial wavenumber and its radial displacement,  $k_r \xi_r$ (i.e., its shear).  We show that $k_r \xi_r \propto \nu_\mathrm{max}^{-9/2}$, implying that the nonlinearity of mixed modes increases significantly as a star evolves.   The modes are weakly nonlinear ($k_r \xi_r > 10^{-3}$) for $\nu_\mathrm{max} \lesssim 150 \, \mu\mathrm{Hz}$ and strongly nonlinear ($k_r \xi_r > 1$) for $\nu_\mathrm{max} \lesssim 30 \, \mu\mathrm{Hz}$, with only a mild dependence on $M$ over the range we consider ($1.0 - 2.0 M_\odot$).   A weakly nonlinear mixed mode can excite secondary waves in the stellar core through the parametric instability, resulting in enhanced, but partial, damping of the mode.  By contrast, a strongly nonlinear mode breaks as it propagates through the core and is  fully damped there.  
Evaluating the impact of nonlinear effects on observables such as mode amplitudes and linewidths requires large mode network simulations. We plan to carry out such calculations in the future and investigate whether nonlinear damping can explain why some red giants exhibit dipole modes with unusually small amplitudes, known as depressed modes.
\vspace{1.1cm}
\end{abstract}

\section{\bf I\lowercase{ntroduction}}
\label{s:intro}

The detection of solar-like oscillations by the \textit{CoRoT} \citep{Baglin:06} and \textit{Kepler} \citep{Borucki:10} space missions has yielded a wealth of information about the internal and global properties of thousands of red giants (see reviews by \citealt{Chaplin:13, Hekker:17}).  Highlights include powerful scaling relations that connect seismic parameters to fundamental stellar parameters (e.g., mass, radii, and luminosity) and the detection of mixed modes, which behave like acoustic waves in the convective envelope and internal gravity waves in the radiative core.  Measurements of mixed mode period spacings make it possible to distinguish between hydrogen- and helium-burning red giants and  \citep{Bedding:11, Mosser:11, Stello:13, Mosser:14} and constrain the core rotation profile \citep{Beck:12,  Deheuvels:12, Deheuvels:14, Mosser:12:rotation}.

The propagation and damping of solar-like oscillations is usually described in terms of the linearized fluid equations. This approximation, in which waves  propagate without interacting, greatly simplifies the analysis of the wave dynamics. In the Sun, acoustic waves ($p$-modes) have sufficiently small amplitude   that the linear approximation is well justified throughout most of the star \citep{CD:02}.  The exceptions are the uppermost  regions of the convective zone and the optically thin region above the photosphere, where the Mach numbers approach one  \citep{Kumar:89}.   However, since there is very little mass in these regions, nonlinear mode interactions do not contribute significantly to the mode damping \citep{Kumar:89} and barely modify the mode frequencies and linewidths \citep{Kumar:94}.

In this paper we argue that, unlike the case for main-sequence stars, nonlinear effects may become important as stars ascend the red giant branch (RGB).  There are two reasons. First, mode amplitudes are observed to increase as stars ascend the RGB, increasing the size of nonlinear effects. Second, for the case of dipole (angular degree $\ell=1$) mixed modes, a new type of nonlinear interaction may become important, namely the steepening of the gravity wave near the center. We investigate the onset of three-wave interactions in the weakly nonlinear limit, as well as the strongly nonlinear limit in which the wave may overturn the stratification near the center, causing the wave to break and deposit its energy there.

Throughout the study, we focus on the stability of low-$\ell$ mixed modes because the observations do not have the spatial resolution to detect modes with $\ell \ga 3$.  We are particularly interested in the stability of pressure-dominated mixed modes ($p$-$m$ modes) since such modes have detectable surface amplitudes and yet propagate deep within the stellar core where nonlinear mode interactions can be important.  

Our calculations rely on RGB models constructed with the \texttt{MESA} stellar evolution code \citep{Paxton:11, Paxton:13, Paxton:15, Paxton:18}.  We consider models with mass $M=[1.0, 2.0] M_\odot$ and $\nu_{\rm max}\simeq [10,200]\muHz$, which coincide with the range observed by \textit{CoRoT} and \textit{Kepler}. We find that the nonlinear mode parameters are not especially sensitive to $M$ and therefore focus on representative models with $M=\{1.2, 1.6, 2.0\}$.  We find eigenmodes of the stellar models with the \texttt{GYRE} oscillation code \citep{Townsend:13, Townsend:18}, and normalize the spatial eigenfunctions $\gv{\xi}_a(\gv{x})$  such that  $\omega_a^2\int d^3x \rho \, |\gv{\xi}_a|^2 = E_\ast$, where $\omega_a$ is the eigenfrequency and $\rho$ is the density.  We express mode energy in units of $E_\ast \equiv GM^2/R$, where $R$ is the stellar radius.
  
The paper is organized as follows.  In Section~\ref{sec:linear_energy}, we estimate the energy of mixed modes as a function of stellar mass $M$ and position on the RGB, or equivalently $\nu_{\rm max}$, the frequency of maximum power.  In Section~\ref{sec:nonlinearity}, we calculate the maximum shear of mixed modes, which provides a measure of their nonlinearity.    In Section~\ref{sec:nonlinear_interactions} we consider the weakly nonlinear regime and study the amplitude equations describing nonlinear three-mode interactions.  We summarize our results in Section~\ref{sec:discussion} and briefly discuss the possibility that the observed depressed modes are due to nonlinear damping.
\section{\bf E\lowercase{nergy of} M\lowercase{ixed} M\lowercase{odes}}
\label{sec:linear_energy}

  By characterizing the power excess of $\simeq 1200$ \textit{Kepler} red giants,  \citet{Mosser:12:depressed}  showed that the bolometric oscillation amplitudes on the RGB are $\sim 10-100$ times larger than the Sun's, and increase dramatically as the stars evolve along the RGB (see also \citealt{Vrard:18}).  The amplitudes are larger because the convective motions are especially vigorous in the low density envelope of red giants (see, e.g., \citealt{Kjeldsen:95, Samadi:07}).  Recent 3D hydrodynamical models are broadly consistent with the observations, and find that the mode excitation rate, $\mathcal{P}$, is a strong function of a star's luminosity-to-mass ratio, scaling as $\mathcal{P} \propto (L/M)^{2.6}$ \citep{Samadi:07, Samadi:12}.  

The time-averaged linear energy of a solar-like oscillation  $E_{a, \rm lin} = \mathcal{P}_a / 2\gamma_a = \mathcal{M}_a v_a^2$, where $\mathcal{P}_a$ is the time-averaged power supplied to the mode by turbulent convection, $\gamma_a$ is the linear damping rate of the mode, $\mathcal{M}_a=MI_a$ is the mode mass, $I_a$ is the dimensionless mode inertia, and $v_a^2$ is the mean-squared surface velocity (see, e.g., \citealt{Belkacem:06}). The linear energy of a $p$-$m$ mode $E_{a, \rm lin} \simeq E_0=\mathcal{P}_0/2\gamma_0$, where $E_0$ is the time-averaged linear energy of the neighboring radial mode ($\ell=0$) of frequency $\nu_0\simeq \nu_a$, with subscript 0 denoting radial modes. The $p$-$m$ mode and radial mode have nearly equal energy because both are damped primarily in the convective envelope, which implies that their work integrals are nearly equal and therefore $\mathcal{M}_a \gamma_a\simeq \mathcal{M}_0 \gamma_0$ \citep{Dupret:09, Grosjean:14}. Moreover, $\mathcal{M}_a \mathcal{P}_a \simeq \mathcal{M}_0\mathcal{P}_0$ because their structures are nearly the same in the convective envelope, where the driving occurs \citep{Dupret:09, Benomar:14, Grosjean:14}.

By fitting the frequency spectra of more than 5000 red giants, \citet{Vrard:18} determine the linewidths $\Gamma_0 = \gamma_0/\pi$ of radial modes with $\nu_0\simeq \nu_{\rm max}$ (see also \citealt{Corsaro:15, Handberg:17}).  They find $\Gamma_{0}(\nu_{\rm max})\approx 0.05-0.2 \muHz$  over the range $M\simeq [0.8, 2.5]M_\odot$ and $\nu_{\rm max}\simeq[10,200]\muHz$.  \citet{Samadi:12} estimate $\mathcal{P}_0(M, \nu_{\rm max})$ from their 3D hydrodynamical models of mode excitation in the upper layers of red giants. They find $\mathcal{P}_0(M,\nu_{\rm max})=B x^s$, where $x = (L/L_\odot)(M_\odot/M)$, $B=4.2^{+1.0}_{-0.8}\times 10^{22} \trm{ erg s}^{-1}$, and $s=2.60\pm 0.08$. Thus,
\bea
&E_{a, \rm lin}(M, \nu_{\rm max})&\simeq \left(1.8^{+0.4}_{-0.3}\right)\times10^{-16}
\non &&
\hspace{-1.8cm}
\times  \left(\frac{\beta}{1.5}\right)^2 \left(\frac{\Gamma_0}{0.1\muHz}\right)^{-1}
\left(\frac{M}{1.5M_\odot}\right)^{-3/2}
\non &&
\hspace{-1.8cm}
\times\left(\frac{T_{\rm eff}}{4800\trm{ K}}\right)^{8.9\pm0.3}\left(\frac{ \nu_{\rm max}}{100\muHz}\right)^{-3.1\mp 0.1}
\label{eq:Ea_obs}
\eea
(in units of $GM^2/R$), where $T_{\rm eff}$ is the effective temperature and we used the scaling relations $L\propto R^2 T_{\rm eff}^4$ and $\nu_{\rm max}\propto M R^{-2}T_{\rm eff}^{-1/2}$  with solar reference values of $\nu_{\rm max,\odot}= 3101\muHz$ and $T_{\rm eff,\odot}=5777\trm{ K}$ \citep{Kjeldsen:95, Stello:09, Huber:10}. We include a correction factor $\beta$ because \citet{Samadi:12} find that the observed bolometric amplitudes $A_0$ are $\approx 1.5$ times larger than those predicted by their hydrodynamical models and suggest that it could be due to their models underestimating $\mathcal{P}_0$ by a factor of $\beta^2$.

We can also express this result in terms of the bolometric amplitude $A_{a, \rm lin} \simeq A_0 \propto \zeta v_0 \propto \zeta (E_0/\mathcal{M}_0)^{1/2}$, where $\zeta$ is a dimensionless coefficient.  \citet{Samadi:12} find $\zeta\simeq (0.59\pm 0.07)x^k$ and $\mathcal{M}_0=Cy^{-p}$, where $y=\Delta \nu/\Delta \nu_\odot$, $\Delta \nu$ is the large frequency separation, $k=0.25\pm 0.05$, $C=8.0^{+2.8}_{-2.1}\times10^{24}\trm{ g}$, and $p=2.0\pm 0.1$. This gives
\bea
&A_{a, \rm lin}(M, \nu_{\rm max})&
{\simeq \left(35^{+11}_{-8}\right) \trm{ ppm}}
\non &&
\hspace{-1.8cm}
\times   \left(\frac{\beta}{1.5}\right) \left(\frac{\Gamma_0}{0.1\muHz}\right)^{-1/2}
\left(\frac{M}{1.5M_\odot}\right)^{-0.25\pm0.01}
\non &&
\hspace{-1.8cm}
\times\left(\frac{T_{\rm eff}}{4800\trm{ K}}\right)^{5.8\pm0.3}\left(\frac{ \nu_{\rm max}}{100\muHz}\right)^{-0.8\pm 0.1},
\label{eq:dLL_obs}
\eea
where we used the relation $\Delta \nu \propto (M/R^3)^{1/2}$ and solar reference values $\Delta \nu_\odot = 134.9\muHz$, $A_{0,\odot} = 2.53\pm 0.11\trm{ ppm}$, and $v_\odot= 18.5\pm 1.5 \trm{ cm s}^{-1}$ \citep{Samadi:12}. Equation (\ref{eq:dLL_obs}) agrees well with the observed bolometric amplitudes measured by \citeauthor{Vrard:18} (2018; see their Figure 6).

\begin{figure}
\centering
\includegraphics[width=3.4in]{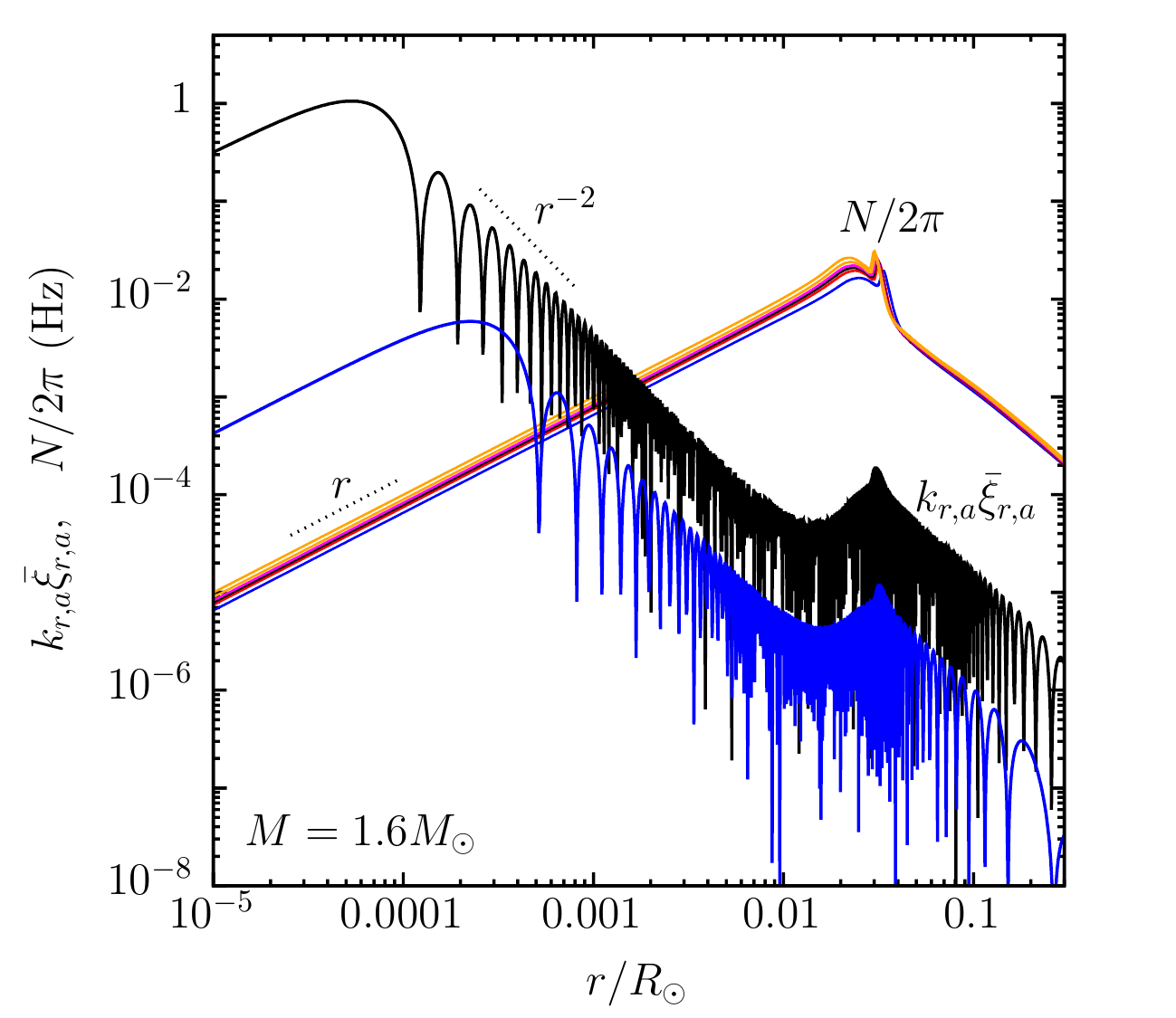} 
\caption{Profiles of $N/2\pi$ and $k_{r,a}\bar{\xi}_{r,a}$ of $p$-$m$ mixed modes (black and blue oscillatory lines) for $M=1.6M_\odot$ RGB models.  The six $N/2\pi$ curves are at different evolutionary stages, corresponding to $\nu_{\rm max}=\{230, 137, 104, 81, 53, 30\}\muHz$ from bottom to top. The black (blue) mixed mode is from the $\nu_{\rm max}=30\muHz$ ($104\muHz$) model, with $\ell_a=1$, $\nu_a\simeq \nu_{\rm max}$ and $E_a$ given by Equation~(\ref{eq:Ea_obs}).
\label{fig:brunt_krxir}}
\end{figure}

\section{\bf N\lowercase{onlinearity of} M\lowercase{ixed} M\lowercase{odes}}
\label{sec:nonlinearity}

A dimensionless local measure of nonlinearity is the shear $d\bar{\xi}_{r,a}/dr\simeq k_{r,a}\bar{\xi}_{r,a}$, where $k_{r,a}$ is the radial wavenumber and $\bar{\xi}_{r,a}=E_a^{1/2}\xi_{r,a}$ is the radial displacement in physical units.  A perturbation at radius $r$ is strongly nonlinear if $k_{r,a}(r)\bar{\xi}_{r,a}(r)\ga 1$, i.e., if the wavelength is smaller than the displacement, since such a wave is likely to overturn and break rather than continue to propagate (similar to ocean waves approaching the shore).  In the core, $p$-$m$ modes are supported by the local buoyancy and  $k_{r,a} \simeq \Lambda_a N/\omega_a r \gg R^{-1}$  for $\nu_a\simeq \nu_{\rm max}$, where $N(r)$ is the Brunt-V\"ais\"al\"a frequency.   Since the wavelengths are small, conservation of WKB flux implies that the radial displacement within the propagation region $\xi_{r, a} \propto r^{-2}$ and the asymptotic eigenmode relations give $|k_r\xi_{r,a}| \approx  K \Lambda_a\omega_a^{-1}r^{-2}$,  where $K= (E_\ast C\Delta P_0 / 2\pi^2\rho)^{1/2}$, $C=N/r$, and $\Delta P_0 = 2\pi^2(\int N d\ln r)^{-1}$  (see, e.g., \citealt{Aerts:10, Hekker:17}).

On the RGB,  $C$ and $K$ are both nearly constant deep within the core.   Figure~\ref{fig:brunt_krxir} shows $N(r)$  for an $M=1.6M_\odot$ model at six different ages, corresponding to $\nu_{\rm max}=\{230, 137, 104, 81, 53, 30\}\muHz$.  Although the core contracts with age, we see that for $r\la0.01R_\odot $, the slope of the $N(r)$ profile is nearly constant with radius, with $C=\{4.1, 4.7, 5.0, 5.3, 5.7, 6.4\} \trm{ s}^{-1} R_\odot^{-1}$, respectively.  Figure~\ref{fig:brunt_krxir} also shows two profiles of $k_{r,a}\bar{\xi}_{r,a}$ for $\ell_a=1$, $\nu_a\simeq \nu_{\rm max}$ $p$-$m$ mixed modes  at $\nu_{\rm max}= 30$ and $104 \muHz$ found with \texttt{GYRE}. The numerical results agree well with the asymptotic expression.  

The shear peaks near the mode's inner turning point $r_a$, which is located where $N(r_a) \simeq C r_a\simeq \omega_a$.  For values characteristic of the RGB models ($E_\star\approx10^{48}\trm{ erg}$, $\rho\approx10^5\trm{ g cm}^{-3}$, $C\approx 5 \trm{ s}^{-1} R_\odot^{-1}$, $\Delta P_0=100\trm{ s}$), $r_a\approx 10^{-4} \nu_{a, 100} R_\odot$ and the maximum shear 
\bea
\left|k_{r,a}\bar{\xi}_{r,a}\right|_{\rm max} &\approx& 0.01 \Lambda_a \left(\frac{\nu_a}{100\muHz}\right)^{-3}\left(\frac{E_a}{10^{-16}}\right)^{1/2}
\non &&\hspace{-1.8cm}
\approx 0.01 \Lambda_a   \left(\frac{\beta}{1.5}\right) \left(\frac{\Gamma_0}{0.1\muHz}\right)^{-1/2} \left(\frac{M}{1.5M_\odot}\right)^{-3/4}
\non &&
\hspace{-1.8cm}
\times\left(\frac{T_{\rm eff}}{4800\trm{ K}}\right)^{9/2}
 \left(\frac{\nu_{\rm max}}{100\muHz}\right)^{-9/2},
\label{eq:krxir}
\eea
where in the second line we plugged in a value of $E_a$ corresponding to the median linear energy $E_{a,\rm lin}$ given by Equation~(\ref{eq:Ea_obs}).   Although the asymptotic eigenmode expressions strictly apply only within the propagation region and not near $r_a$, they approximate the magnitude and scaling of $\left|k_{r,a}\bar{\xi}_{r,a}\right|_{\rm max}$ very well.

\begin{figure}
\centering
\includegraphics[width=3.4in]{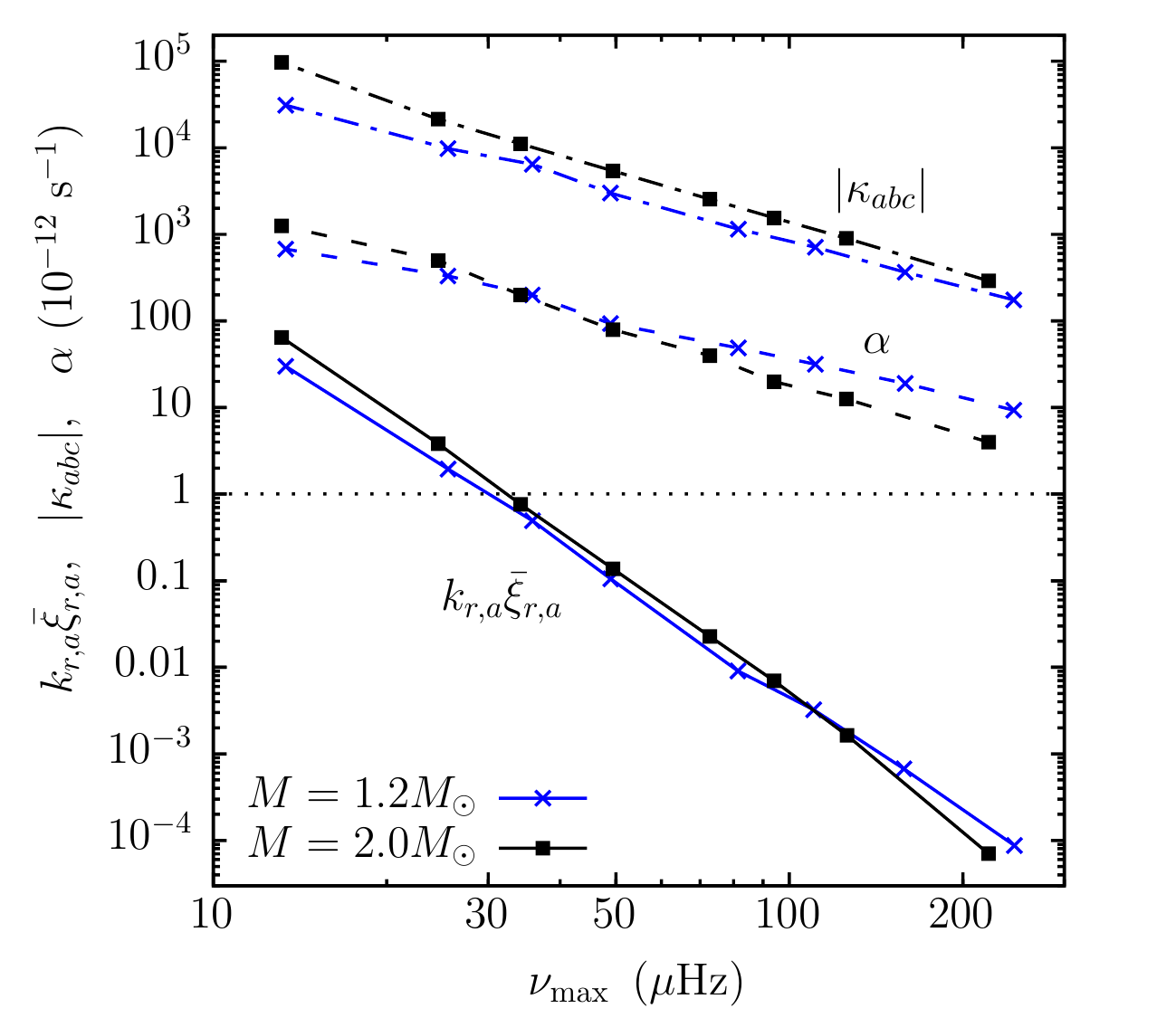} 
\caption{
Maximum shear $k_{r,a}\bar{\xi}_{r,a}$ of $p$-$m$ mixed modes found using \texttt{GYRE} (solid lines)
as a function of $\nu_{\rm max}$ for $M=1.2 M_\odot$ (blue curves with crosses) and $2.0 M_\odot$ (black curves with squares). Also shown are the nonlinear coupling coefficient $\kappa_{abc}$ (dashed-dotted lines), and linear damping coefficient $\alpha$ (dashed lines; in units of $10^{-12}\trm{ s}^{-1}$). The modes are $(\ell_a, \ell_b,\ell_c)=(1,2,3)$ with $\nu_a\simeq \nu_{\rm max}$ and $\nu_b+\nu_c\simeq \nu_a$.
\label{fig:gamma_kappa}}
\end{figure}

Since the linear energy of  mixed modes near $\nu_{\rm max}$ scales approximately as $E_{a,\rm lin}\propto \nu_a^{-3}$, the maximum shear increases significantly as the star evolves ($|k_{r,a}\bar{\xi}_{r,a}|_{\rm max} \propto \nu_a^{-9/2}$). Figure~\ref{fig:gamma_kappa} shows the numerically calculated maximum shear as a function of $\nu_{\rm max}$ for $p$-$m$ modes of the $M=1.2 M_\odot$ and $2.0 M_\odot$ models.  To calculate the maximum shear, we use \texttt{GYRE} to find modes with $\ell_a=1$, $\nu_a\simeq \nu_{\rm max}$ and from these numerical solutions we compute the maximum of $d\xi_{r,a}/dr \simeq k_{r,a}\xi_{r,a}$ for each mode.  We then use Equation~(\ref{eq:Ea_obs})  to calculate $E_{a, \rm lin}$.  We take $\beta=1.5$, $\Gamma_0=0.1\muHz$, and the median values for $B$ and $s$.  The analytic expression for $|k_{r,a}\bar{\xi}_{r,a}|_{\rm max}$ given by Equation (\ref{eq:krxir}) agrees with the numerical result to within a factor of $\approx 2$ over the range of $M$ and $\nu_{\rm max}$ shown in Figure~\ref{fig:gamma_kappa}.  We find that for $50\la \nu_{\rm max}\la 150\muHz$, the mixed modes are weakly nonlinear ($0.1\ga k_{r,a}\bar{\xi}_{r,a} \ga 10^{-3} $).  However, for $\nu_{\rm max}\la 30\muHz$ they become strongly nonlinear in the core ($k_{r,a}\bar{\xi}_{r,a} \ga 1$).  

Despite the uncertainties in $E_{a,\rm lin}$ (due to uncertainties and observational scatter in the parameters that determine the energy), the steep $\nu_{\rm max}^{-9/2}$  dependence implies that there is a narrow  $\nu_{\rm max}$ window where the modes transition from weakly nonlinear to strongly nonlinear.  A strongly nonlinear mixed mode will overturn the local stratification in the core and break.  Since they do not reflect at $r_a$, they are ingoing traveling waves rather than standing waves.    This phenomenon can also occur in the context of dynamical tides, where it can lead to rapid, tide-induced orbital evolution (see, e.g., \citealt{Goodman:98, Barker:10, Weinberg:17}).  In the present context, the breaking wave is directly observed.

 As a star evolves, radiative damping in the core becomes so strong that it can dissipate all the energy from a mode in less than its group travel time across the star.    For gravity-dominated mixed modes ($g$-$m$ modes), this transition occurs at $\nu_{\rm max}\approx 30\muHz$ \citep{Dupret:09, Grosjean:14}, similar to where wave breaking occurs.  However, for $p$-$m$ modes a significant portion of their energy remains trapped in the envelope and is not lost to radiative damping.

\section{\bf N\lowercase{onlinear} M\lowercase{ode} I\lowercase{nteractions}}
\label{sec:nonlinear_interactions}

In the previous section we found that mixed modes are weakly nonlinear over a broad portion of the lower RGB ($\nu_{\rm max}\approx [50, 200]\muHz$), before becoming strongly nonlinear as the star evolves up the RGB.  Weakly nonlinear waves can excite secondary waves through nonlinear mode interactions. Whether a weakly nonlinear primary wave (a parent mode) is unstable to secondary waves (daughter modes) depends on the parent's amplitude, the strength of the nonlinear interactions, and the damping rates and frequency detunings.  In this section we assess the stability of weakly nonlinear mixed modes as a function of $M$ and $\nu_{\rm max}$.  We focus on the stability of $\ell=1$ $p$-$m$ parent modes with $\nu_a\simeq \nu_{\rm max}$ coupled to resonant $g$-$m$ daughter modes.

To account for weakly nonlinear effects,  we expand the Lagrangian displacement field as a sum of linear eigenmodes $\gv{\xi}(\gv{x},t)=\sum q_a(t) \gv{\xi}_a(\gv{x})$ and keep terms up to $\mathcal{O}\left(\gv{\xi}^2\right)$.  The equation of motion for $\gv{\xi}(\gv{x},t)$ can then be written a set of coupled, nonlinear amplitude equations \citep{Dziembowski:82,Kumar:96,Wu:01, Schenk:02, Weinberg:12},
\bea
\ddot{q}_a + 2\gamma_a \dot{q}_a + \omega_a^2 q_a &=& \omega_a^2 f_a(t) + \omega_a^2 \sum_{b}\sum_c \kappa_{abc}^\ast q_b^\ast q_c^\ast,
\hspace{0.2cm}
\label{eq:modeampeqn}
\eea
where $\omega_a$ and $\gamma_a$ are the eigenfrequency and linear damping rate of mode $a$ and the asterisks denote complex conjugation.   The linear forcing $f_a(t)$ accounts for the stochastic excitation of mode $a$ due to turbulent motions at the top of the convective envelope.  The sum containing the dimensionless three-mode coupling coefficient $\kappa_{abc}$ accounts for the nonlinear interaction between mode $a$ and other modes $b, c$. The modes couple only if they satisfy the angular selection rules $|\ell_b-\ell_c|\le \ell_a \le \ell_b+\ell_c$ with $\ell_a+\ell_b+\ell_c$ even and $m_a+m_b+m_c=0$ ($\ell$ is the angular degree and $m$ is the azimuthal order). 

We study the stability of weakly nonlinear mixed modes by analyzing Equation~(\ref{eq:modeampeqn}) for simple three-mode systems.  In Section~\ref{sec:linear_forcing} we describe our treatment of the linear stochastic forcing $f_a(t)$.  In Section~\ref{sec:stability_analytic} we present analytic estimates of the stability criterion and daughter growth rates and show example numerical solutions of Equation~(\ref{eq:modeampeqn}).  In Section~\ref{sec:nonlinear_parameters} we evaluate the various mode parameters that enter the stability analysis.  In Section~\ref{sec:Eth} we use these results to determine the stability of mixed modes  as a function of $M$ and $\nu_{\rm max}$.

\subsection{Linear Stochastic Forcing}
\label{sec:linear_forcing}

The modes are excited by the large number of granules in the upper regions of the convection zone, which each impart  small, independent impulses.  The most strongly excited modes are those with periods comparable to the eddy turnover time $\omega_a^{-1} \sim \tau_{\rm eddy}$ \citep{Goldreich:88}.  Since the size of each granule is of order a scale height $H$, there are approximately $(R/H)^2 \gg 1$ granules, and the mean time between impulses $\Delta t \sim \tau_{\rm eddy} (H/R)^2 \ll \omega_a^{-1}$.  \citet{Chang:98} estimate that in the Sun $(R/H)^2\sim 10^5$ and $\tau_{\rm eddy} \sim  15\trm{ min}$, which implies $\Delta t \sim 10^{-3}\trm{ s}$, i.e., about $10^5$ impulses per mode period.  In red giants, the impulse rate can be even larger.  By contrast, the damping rate of the mode is on a much longer timescale, $\gamma_a^{-1} \gg \tau_{\rm eddy}$.  

Similar to previous studies (e.g., \citealt{Kumar:88, Chang:98}), we model the stochastic forcing as a Poisson process involving a random sequence of  impulses at times $t_j$.  We assume that the time between consecutive  impulses, $\Delta t=t_{j+1}-t_j$, is an independent random variable whose probability density is given by $p(\Delta t) = \mu \exp(-\mu \Delta t)$, where $\mu$ is the mean number of impulses per unit time.  The mode forcing $f_a(t)$ is the sum of all the individual impulses
\beq
f_a(t) = f_{0,a} \sum_j c_j \psi_j(t),
\eeq
where we assume each impulse has a Gaussian time dependence $\psi_j(t)=\exp[-(t-t_j)^2/2\tau^2]$. Since we expect the correlation time of each impulse $\tau \sim \tau_{\rm eddy}$, we set $\omega_a\tau=1$.  The amplitude $c_j = c_{j,r}+ic_{j,i}$ is complex with $c_{j,r}$ and $c_{j,i}$ drawn from a Gaussian probability distribution centered on zero with standard deviation one.  The constant $f_{0,a}$ sets the overall scale of the mode amplitude. While the discussion above suggests $\mu > 10^5$ per mode period, we find that our numerical results are insensitive to $\mu$ as long as we set $\mu \ga 10$ per mode period.

\begin{figure}
\centering
 \includegraphics[width=3.4in]{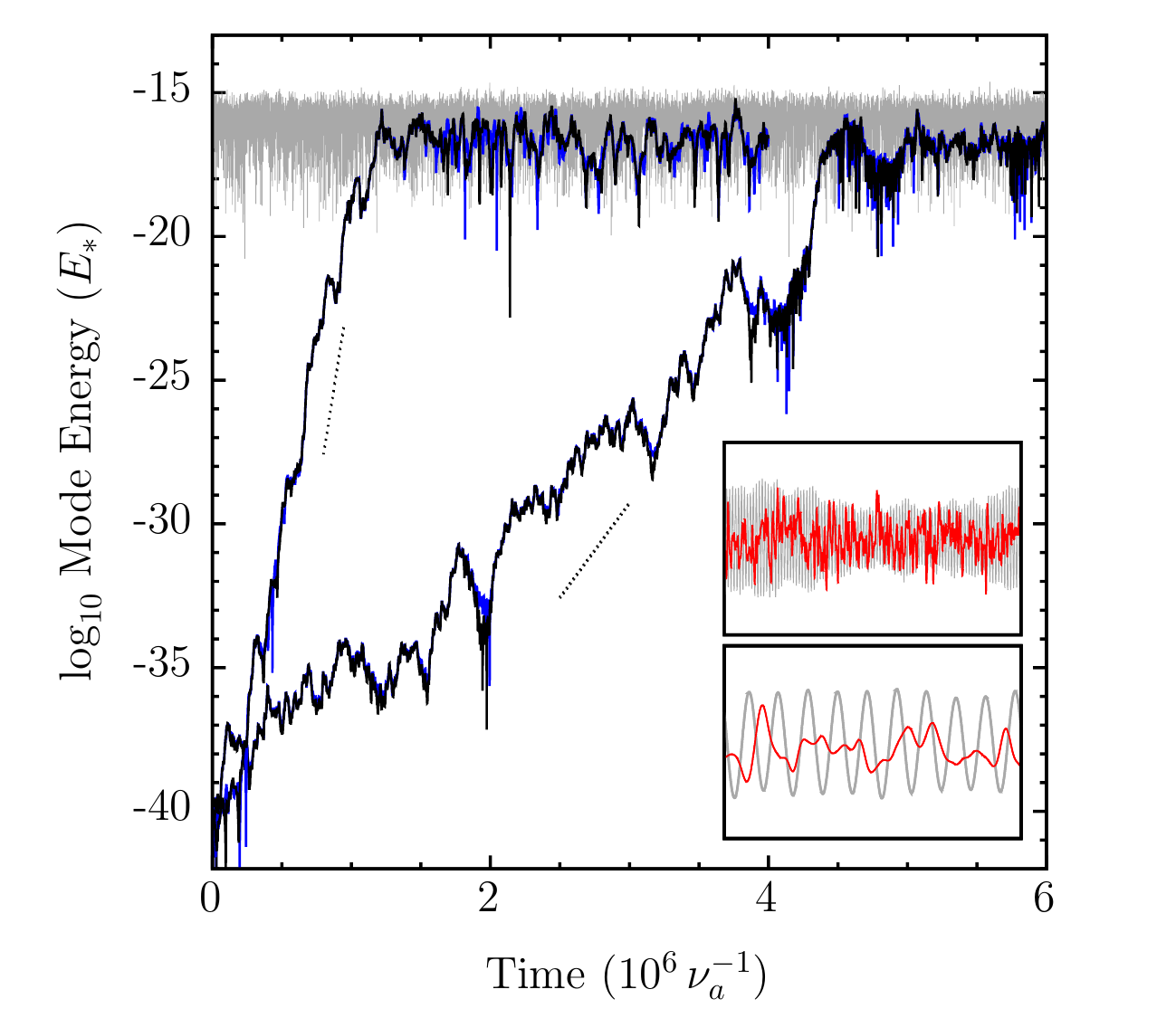} 
\caption{Mode energy as a function of time (main panel) for two examples of a three-mode system involving a stochastically driven parent (gray line) coupled to a resonant daughter pair (black and blue lines) with $f_b=f_c=0$ and initial energies of $E_b\simeq E_c=10^{-40}$.  The examples differ only by $\gamma_a =(\pi, 2\pi) \times10^{-4}\nu_a$, with the smaller $\gamma_a$ corresponding to the more rapidly growing daughter pair.  The other parameters are $\mu = 100 \nu_a$, $E_{a, \rm lin} \simeq 2\times10^{-16}$, $\gamma_b\simeq\gamma_c\simeq10^{-5}\nu_a$, $\Delta_b\simeq\Delta_c\simeq 10^{-5} \omega_a$, and $\kappa_{abc}=3\times10^3$. The dashed lines show the growth rate scalings given by Equation (\ref{eq:growth_rate}).  The insets zoom in on the parent mode and show, on a linear scale, the real parts of $q_a(t)$ (gray lines) and $10 f_a(t)$ (red lines) over durations of $10\nu_a^{-1}$ and $100\nu_a^{-1}$. 
\label{fig:amp_eqn}}
\end{figure}

One realization of this random process corresponds to a set of $t_j$ and $c_j$. Each realization, $\mathcal{R}$, will produce a different solution, $q_a^{(\mathcal{R})}(t)$, to Equation (\ref{eq:modeampeqn}) for the parent amplitude. The ensemble average, which we denote by angle brackets, corresponds to the average of all these realizations. This ensemble average can be carried out either by averaging many numerical simulations, or in analytic work by directly averaging over the Poisson distribution for the $t_j$ and the Gaussian distribution for the $c_j$. In addition, single realizations that include very large numbers of events are expected to approximate the ensemble average. Hence even though single realization results are shown in Figure~\ref{fig:amp_eqn}, over long timescales we expect the results to be comparable to ensemble averaging.

In the absence of nonlinear coupling, $q_a(t)$ satisfies the equation of a damped linear oscillator forced by a stationary random function for which different events are uncorrelated. The ensemble average of the parent energy, which we denote as $E_{a,\rm lin}$, is
\beq
E_{a,\rm lin} \equiv \langle |q_a(t)|^2 \rangle  \approx  (f_{0,a} \omega_a \tau)^2  \left( \frac{\mu}{2\gamma_a } \right).
\label{eq:Ealin}
\eeq
This expression can be understood as follows. Integrating the forcing $ \propto f_{0,a} \omega_a$ over the impulse time $\tau$ gives an amplitude of $f_{0,a} \omega_a \tau$ for one impulse. Over the damping time $\gamma_a^{-1}$, there are $\mu/\gamma_a$ impulses which add randomly, giving the result in Equation (\ref{eq:Ealin}).

Ensemble averaging products of first order amplitudes requires the autocorrelation function of the forcing. For the daughters, the random forcing function involves the parent amplitude. For our model, the correlation function is approximately
\beq
\langle q_a^*(t) q_a(t') \rangle \simeq E_{a,\rm lin} e^{-\gamma_a |t-t'|} \cos \left[ \omega_a (t-t') \right].
\eeq
This correlation function oscillates at the parent frequency, and has an exponential dependence with correlation time $\gamma_a^{-1}$, the damping time of the parent. The growth rate of the daughters relies on the Fourier transform of this correlation function. Define the power $P_a(\omega)$ to be
\begin{eqnarray}
P_a(\omega) & \equiv & {\rm Re} \int_0^\infty d(t-t') e^{-i\omega(t-t')} \langle q_a^*(t) q_a(t') \rangle
\nonumber \\ & \simeq & \frac{E_{a,\rm lin}}{2}  \int_0^\infty du   \cos[(\omega-\omega_a) u] e^{ - \gamma_a u}
\nonumber \\ & \simeq & \frac{E_{a,\rm lin}}{2}  \left[ \frac{\gamma_a}{(\omega-\omega_a)^2 + \gamma_a^2} \right].
\end{eqnarray}
Hence the power is a Lorentzian with damping $\gamma_a$ and detuning $\omega-\omega_a$  (see, e.g., \citealt{CD:89}).  We will use this result when describing the nonlinear stability of parent modes in Section~\ref{sec:stability_analytic}

The two insets in Figure~\ref{fig:amp_eqn} show the real part of $q_a(t)$ and $10\times f_a(t)$ over a duration of $10 \nu_a^{-1}$ and $100\nu_a^{-1}$, where $\nu_a=\omega_a/2\pi$.  On such timescales, $q_a(t)$ looks like a sinusoidal oscillation with a slowly varying amplitude.  The force varies stochastically on a timescale $\approx \nu_a^{-1}$ and has characteristic strength of $|f_a| \ll |q_a|$. 

\subsection{Nonlinear Stability}
\label{sec:stability_analytic}

Consider a stochastically driven parent mode with $\nu_a\simeq \nu_{\rm max}$ that is coupled to a single, self-coupled daughter mode with a frequency of $\nu_b\simeq \nu_a/2$. Observationally, the distribution of mode energies can be fit by a Gaussian envelope with full-width at half-maximum   $\delta \nu_{\rm env} \simeq 0.66 (\nu_{\rm max}/100\muHz)^{0.88}$ \citep{Mosser:12:depressed}.  A daughter mode with a frequency of $\nu_b \simeq \nu_a/2 \simeq \nu_{\rm max}/2$ therefore has a linear energy that is only $\sim 1\%$ that of its parent.  Thus, to a good approximation we can ignore the daughter's linear forcing and set $f_b=0$. The amplitude equation then reduces to the stochastic Mathieu equation (when $|q_b| \ll |q_a|$), whose stability has been studied extensively (see, e.g., \citealt{Stratonovich:65, Ariaratnam:76, vanKampen:92, Zhang:93, Poulin:08}).  In the regime relevant to mixed modes in red giants, it can be shown that the daughter's average nonlinear growth rate 
$s_b \simeq -\gamma_b +2\omega_b^2\kappa_{abb}^2P_a(2\omega_b)$, 
i.e., 
\beq
s_b \simeq -\gamma_b + \frac{\omega_b^2 \kappa_{abb}^2 E_{a, \rm lin}}{\gamma_a} \left[1+\frac{\Delta_b^2}{\gamma_a^2}\right]^{-1},
\label{eq:growth_rate}
\eeq
where $\Delta_b = \omega_a-2\omega_b$ is the daughter detuning and we assume $\gamma_a \gg \gamma_{b,c}$ (see Sections~\ref{sec:gamma_a} and~\ref{sec:gamma_b}).  Thus, the parent and daughter are ``parametrically unstable" ($s_b>0)$ if the parent linear energy $E_{a, \rm lin}$ is larger than a threshold energy 
\beq
E_{\rm th} = \frac{\gamma_a \gamma_b}{\kappa_{abb}^2\omega_b^2}\left[1+\frac{\Delta_b^2}{\gamma_a^2}\right].
\label{eq:Eth}
\eeq
For characteristic parameter values (see Section~\ref{sec:nonlinear_parameters})
\bea
E_{\rm th} &\simeq &2.5 \times 10^{-16} \left(\frac{\gamma_a}{10^{-7} \trm{ s}^{-1}}\right)\left(\frac{\gamma_b}{10^{-9} \trm{ s}^{-1}}\right)
\non &&\times\left(\frac{\kappa_{abb}}{10^3}\right)^{-2} \left(\frac{\nu_b}{100 \muHz}\right)^{-2}\left[1+\frac{\Delta_b^2}{\gamma_a^2}\right].
\eea
To understand the scaling with $\gamma_a$, note that at a given $E_{a, \rm lin}$,  $P_a(2\omega_b)\propto \gamma_a^{-1}$ provided that $|\Delta_b|\ll\gamma_a$.  Thus, as $\gamma_a$ increases, the parent's power is less concentrated near $\omega_a\simeq 2\omega_b$ and the daughter driving is  less effective, resulting in smaller $s_b$ and larger $E_{\rm th}$.  

Our estimates above assume that only a single daughter pair ($N=2$) is parametrically excited.  However, in studying tidal flows, \citet{Weinberg:12} found that sets of $N\gg2$ daughters can be collectively excited (see also \citealt{Essick:16}), and that that the growth rate of collective sets is larger by a factor of $N$ (and $E_{\rm th}$ is smaller by a factor of $N$).  If mixed mode parents excite collective sets of unstable daughters, then  $s_b$ ($E_{\rm th}$) can be significantly larger (smaller) than the above estimates.

In Figure~\ref{fig:amp_eqn} we show two examples of parametrically unstable three-mode systems in which a stochastically driven parent is coupled to a resonant daughter pair with $f_b=f_c=0$.  The examples differ only in the assumed value of $\gamma_a$. Although there are two daughter modes rather than a single self-coupled daughter, the daughters are  similar ($\gamma_b\simeq\gamma_c$, $\omega_b\simeq \omega_c$).  We find that they grow in a stochastic fashion and have an instability threshold and average growth rate that agrees reasonably well with Equations~(\ref{eq:growth_rate}) and (\ref{eq:Eth}).  The stochastic nature of the driving necessarily implies that the growth rates vary rapidly with time, and numerically we find that different realizations only approach the ensemble average over long timescales.

If the parent is driven harmonically rather than stochastically, the daughters satisfy the standard Mathieu equation.  They would then be subject to the usual parametric subharmonic instability (PSI), with $s_b \simeq -\gamma_b + 2\omega_b \kappa_{abb} E_{a, \rm lin}^{1/2}$ and $E_{\rm th} \simeq (\gamma_b/2\kappa_{abb}\omega_b)^2$, assuming $|\Delta_b| \ll \gamma_b$ (see, e.g., \citealt{Dziembowski:82, Wu:01}).  For parameter values relevant to the coupling of mixed modes on the RGB,  the stochastic growth rate is smaller than the PSI rate by a factor of $\omega_b \kappa_{abb} E_{a, \rm lin}^{1/2}/2\gamma_a \ll 1$, and the stochastic energy threshold is larger by a factor of $4\gamma_a/\gamma_b \gg 1$.  In numerical experiments, we find that if we choose (artificial) parameter values such that $\omega_b \kappa_{abb} E_{a, \rm lin}^{1/2}/2\gamma_a \ga 1$, then the daughter grows at the PSI rate rather than the stochastic rate (the latter now being the larger of the two rates). Indeed, for small enough $\gamma_a$ we expect to recover the PSI since $P_a(\omega_a)$ is so narrowly peaked that, as far as the resonant daughters are concerned, the parent oscillates harmonically.

\subsection{Mode Parameters}
\label{sec:nonlinear_parameters}

\subsubsection{$\gamma_a$}
\label{sec:gamma_a}

As described in Section~\ref{sec:linear_energy}, $\gamma_a \simeq \mathcal{M}_0 \gamma_0 / \mathcal{M}_a$ and $\Gamma_{0}=\gamma_0/\pi \approx 0.05-0.2 \muHz$.  For $\ell_a=1$ $p$-$m$ modes, we find using \texttt{GYRE} that the inertia ratio is $\mathcal{M}_0/\mathcal{M}_a \simeq 0.1-0.5$.  The exact value depends on how close a particular $p$-$m$ mode is to an acoustic cavity resonance.  Those closest to a resonance are the most $p$-mode-like and have $\mathcal{M}_0/\mathcal{M}_a \simeq 0.3-0.5$, while those $p$-$m$ modes on either side of a resonance have $\mathcal{M}_0/\mathcal{M}_a \simeq 0.1-0.3$   \citep{Goupil:13, Deheuvels:15, Mosser:15}.  Thus, we estimate that $10^{-8} \la \gamma_a  \la 10^{-7}\trm{ s}^{-1}$, which agrees well with the available measurements of individual $p$-$m$ mode linewidths \citep{Mosser:18}. 

\subsubsection{$\gamma_{b,c}$}
\label{sec:gamma_b}

Since daughter $g$-$m$ modes with $\ell_b \ga 2$ are well trapped in the core, they undergo radiative damping in the core but comparatively  little damping in the convective envelope.  As a result, they tend to have much smaller damping rates than $p$-$m$ modes as long as $\nu_{\rm max} \ga 30 \muHz$ \citep{Dupret:09, Grosjean:14, Mosser:18}.  

We can estimate the contribution of the convective envelope to the damping by computing  $\mathcal{M}_0 \gamma_0 / \mathcal{M}_b$ as in Section~\ref{sec:gamma_a}.  For $\ell_b =2$ $g$-$m$ modes, we find $\mathcal{M}_0/\mathcal{M}_b \la 10^{-3}$ at $\nu_{\rm max}\simeq 200\muHz$ for  $M=1.2M_\odot$ and $M=2.0 M_\odot$; the  inertia ratio is even smaller for larger $\ell_b$ and at smaller $\nu_{\rm max}$ because as the star evolves, the core contracts and the $g$-$m$ modes become even more strongly trapped in the $g$-mode cavity.  Thus, for modes with $\ell_b\ge 2$, the convective envelope contributes $\la 5\times 10^{-10}\trm{ s}^{-1}$ given that $\gamma_0/\pi \approx 0.05-0.2 \muHz$.   As we now describe, this is smaller than the contribution from  radiative damping in the core.

Using the non-adiabatic calculations in \texttt{GYRE}, which only account for radiative damping, we find $\gamma_b\approx \alpha \Lambda_b^2 \nu_{b,100}^{-2}$, where $\Lambda_b^2 = \ell_b(\ell_b+1)$, $\nu_{b,100} = \nu_b / 100 \muHz$, and $\alpha$ is a model-dependent constant.  The quadratic scaling is a consequence of the short-wavelength of the modes (see \citealt{Hekker:17}). Values of $\alpha(M, \nu_{\rm max})$ for $M=1.2 M_\odot$ and $2.0 M_\odot$ are shown in Figure~\ref{fig:gamma_kappa}.  We find $0.05\la \alpha / 10^{-10}\trm{ s}^{-1} \la 10$  over the $M$ and $\nu_{\rm max}$ range of our models. There is a strong dependence on $\nu_{\rm max}$ because as the star evolves, the core contracts and $N$ in the core increases (see Equation (3) in \citealt{Dupret:09}). The dependence on $M$ is fairly weak and non-monotonic ($\alpha$ increases for $M\la 1.6 M_\odot$ and then decreases, like the core density).

For example, a resonant daughter pair with $(\ell_b,\ell_c)=(2,3)$  coupled to a parent with $\nu_a\simeq\nu_{\rm max}=100\muHz$ has $(\gamma_b,\gamma_c)\simeq(2, 5)\times10^{-9}\trm{ s}^{-1}$ (since $\alpha\simeq 10^{-10}\trm{ s}^{-1}$).  For comparison, \citet{Grosjean:14}, who  account for damping in both the core and envelope,  find that their models~$\{\trm{E, B, F, G}\}$  ($M=\{1.0, 1.5, 1.7, 2.1\} M_\odot$ and $\nu_{\rm max}=\{88, 97, 90, 66\}\muHz$) all yield lifetimes of $\gamma^{-1}\simeq 2000\trm{ days}$ (i.e., $\gamma\simeq6\times10^{-9}\trm{ s}^{-1}$) for $\ell=2$ modes with $\nu\simeq\nu_{\rm max}$.   They do not show results for modes with $\ell>2$, but their Figure 4 suggests that such modes might have smaller $\gamma$ than their $\ell=2$ modes (since they are even more strongly trapped in the core). Note too that their calculations seem to overestimate the damping rates of radial modes by a factor of $\sim 10$ (they find lifetimes of $\gamma_0^{-1}\approx 3\trm{ days}$ whereas the observations by \citet{Vrard:18} suggest $\gamma_0^{-1}\ga 20\trm{ days}$, i.e., $\Gamma_0\la 0.2 \muHz$).

\subsubsection{$\Delta_{b,c}$}
\label{sec:Dbc}

The minimum daughter detuning is $|\Delta_b|  \approx  \omega_a \ell_b^{-3} n_a^{-2}$, where $n_a$ is the radial order of the parent \citep{Wu:01}.  One factor of $n_a$ comes from the mean period spacing of mixed modes and the other comes from the number of well-coupled daughters given the width of maximum $|\kappa_{abc}|$ (daughters with $|n_b-n_c|\la n_a$ all have similar $\kappa_{abc}$; \citealt{Kumar:96, Weinberg:12}).  The $\ell_b^{-3}$ dependence (or $\ell_b^{-2}$ if rotation does not lift the $m$ degeneracy) comes from the freedom in choosing daughters allowed by the angular selection rules. Mixed modes near $\nu_{\rm max}$ have short wavelengths in the core and $n_a \approx \Lambda_a/\nu_a \Delta P_0$  \citep{Hekker:17}.  We find $90\la \Delta P_0 \la 130 \trm{ s}$ for $M=[1.0,1.8]M_\odot $ and $\nu_{\rm max}=[50,200]\muHz$ (larger $\nu_{\rm max}$ and smaller $M$ have larger $\Delta P_0$). This implies that for $\ell_a=1$, $\ell_b=2$, and $\nu_{\rm max} \la 100\muHz$, $|\Delta_b|  \la  10^{-5}\omega_a$, which agrees well with eigenmode searches with \texttt{GYRE}.  Given the $\gamma_a$ estimate above,    $|\Delta_b|< \gamma_a$ for $\nu_{\rm max} \la 100\muHz$. Thus, the more evolved models always have daughters with sufficiently small $|\Delta_b|$ that detuning does not limit their growth rate or $E_{\rm th}$  (see Equations~(\ref{eq:growth_rate}) and (\ref{eq:Eth})). 

\begin{figure}
\centering
\hspace{-0.7cm}
\includegraphics[width=3.1in]{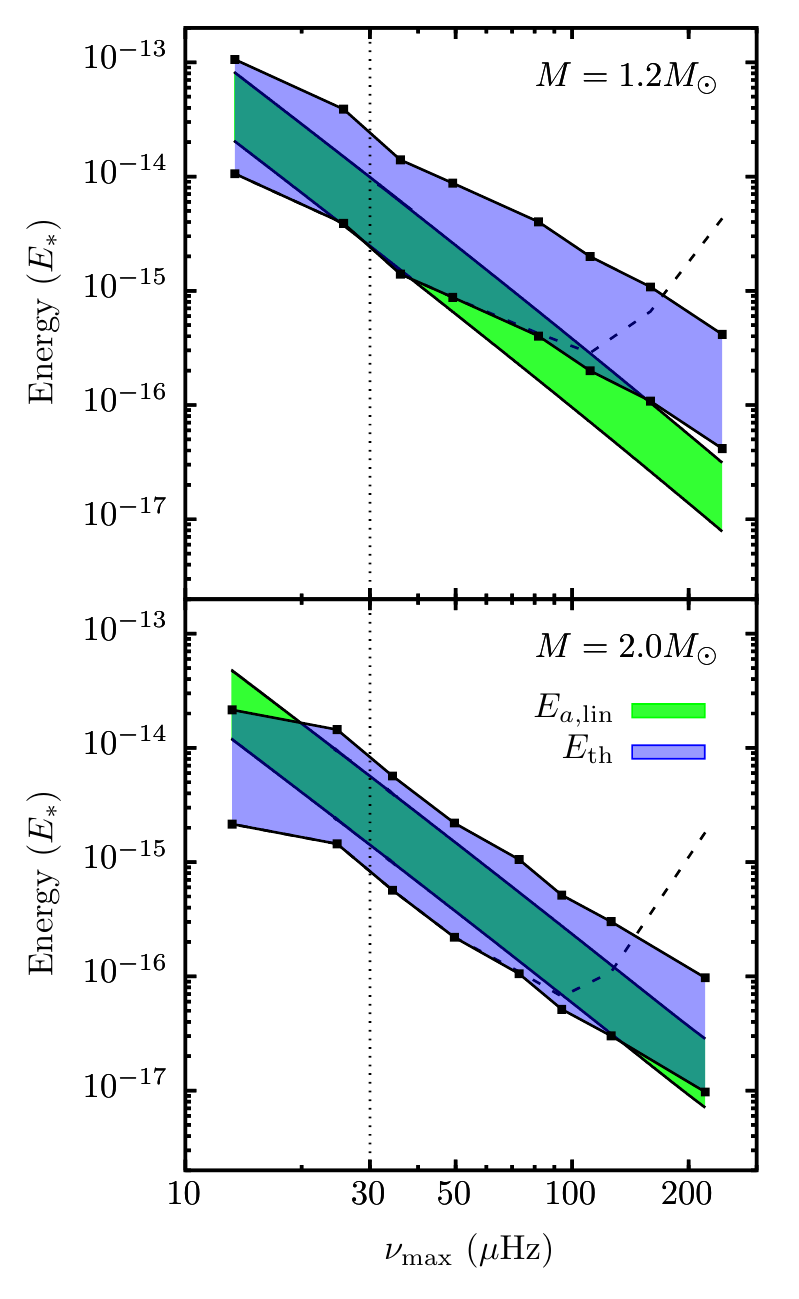} 
\caption{Linear energy $E_{a,\rm lin}$ (green regions) and nonlinear threshold energy $E_{\rm th}$ (blue regions) as a function of $\nu_{\rm max}$ for $M=1.2 M_\odot$ (top panel) and $M=2.0 M_\odot$ (bottom panel). The factor of 4 range in $E_{a, \rm lin}$ at a given $\nu_{\rm max}$ reflects the observed range of $\Gamma_0$; the lower and upper envelopes correspond to $\Gamma_0= 0.2\muHz$ and $0.05\muHz$, respectively.  The factor of 10 range in $E_{\rm th}$ reflects the range of possible $\gamma_a$; the lower and upper envelopes correspond to $\gamma_a= 10^{-8}\trm{ s}^{-1}$ and $10^{-7}\trm{ s}^{-1}$, respectively.  The $E_{\rm th}$ curves assume $(\ell_a,\ell_b,\ell_c)=(1,2,3)$, $\nu_a\simeq\nu_{\rm max}$, and fortuitous detuning $|\Delta_b|\ll \gamma_a$.  The dashed curve shows a portion of $E_{\rm th}$ assuming instead that $|\Delta_b|=\omega_a\ell_b^{-3} n_a^{-2}$. The vertical dotted line indicates the approximate $\nu_{\rm max}$ where mode $a$ becomes strongly nonlinear ($k_{r,a}\bar{\xi}_{r,a} > 1$).
\label{fig:Eth_Ealin}}
\end{figure}

\subsubsection{$\kappa_{abc}$}
\label{sec:kappa}

In order to calculate $\kappa_{abc}(M,\nu_{\rm max})$, we search for  eigenmode triplets with \texttt{GYRE} and use the expression for $\kappa_{abc}$ given in \citeauthor{Weinberg:12} (2012; see their (A55)-(A62)). The coupling occurs primarily near the inner turning radius of the parent, deep in the stellar core, since that is where the parent's shear $k_{r,a}\xi_{r,a}$ peaks (see Section~\ref{sec:nonlinearity}). Daughters with similar wavenumber $|k_{r,b}-k_{r,c}| \la  k_{r,a}$ are spatially resonant with the parent and therefore couple strongly to it. Since deep in the core $N \propto r$, good spatial resonance and small detuning imply $ \omega_b/\omega_a\simeq \Lambda_b/(\Lambda_b+\Lambda_c)$, and similarly for $\omega_c$.  For a given $\ell_a$, we use this condition and the angular selection rules to find resonant daughters that maximize  $|\kappa_{abc}|$.

Figure~\ref{fig:gamma_kappa} shows the maximum $|\kappa_{abc}|$ for $M=1.2M_\odot$ and $M=2.0M_\odot$ as a function of $\nu_{\rm max}$ assuming $\nu_a\simeq \nu_{\rm max}$, resonant daughters, and $(\ell_a,\ell_b,\ell_c)=(1,2,3)$.  We find $\kappa_{abc}\simeq \kappa_0 \nu_{a,100}^{-2}$, where $\kappa_0\simeq \{900, 1100, 1400\}$ for $M=\{1.2, 1.6, 2.0\} M_\odot$.  
To understand the magnitude of $\kappa_0$ and the $\nu_{\rm max}^{-2}$ scaling, note that while the exact expression for $\kappa_{abc}$ is complicated and contains many terms,  \citeauthor{Weinberg:12} (2012; see their Equation (43)) showed that the dominant terms scale with the parent shear and imply 
$\kappa_{abc}\approx (T\Delta P_0 /2\pi^2) \int N k_{r,a}\xi_{r,a} d\ln r$,  where the angular integral $|T|\approx 0.2$ for low-degree modes.  Using the asymptotic relation for $k_{r,a} \xi_{r,a}$ given in Section~\ref{sec:nonlinearity}, we find $\kappa_{abc} \approx  (\Lambda_a TKC^2 \Delta P_0 /8\pi^4) \nu_a^{-2}$. Plugging in characteristic values from the stellar models ($E_\star\approx10^{48}\trm{ erg}$, $\rho\approx10^5\trm{ g cm}^{-3}$, $C\approx 5 \trm{ s}^{-1} R_\odot^{-1}$, $\Delta P_0=100\trm{ s}$) gives   $\kappa_{abc}\approx 10^3  \nu_{a,100}^{-2}$ for $(\ell_a,\ell_b,\ell_c)=(1,2,3)$, in good agreement with the full $\kappa_{abc}$ calculation.

\subsection{Nonlinear Energy Threshold}
\label{sec:Eth}
 
 From the estimates of the various mode parameters given in Section~\ref{sec:nonlinear_parameters}, we can calculate the nonlinear energy threshold, $E_{\rm th}(M,\nu_{\rm max})$  and compare it to $E_{a,\rm lin}(M,\nu_{\rm max})$ (see  Equations~(\ref{eq:Eth}) and (\ref{eq:Ea_obs})). Representative results are shown in Figure~\ref{fig:Eth_Ealin} for $M=1.2M_\odot$ and $M=2.0M_\odot$ assuming a mixed mode parent with $\ell_a=1$ and $\nu_a\simeq \nu_{\rm max}$. The green region shows the possible range of $E_{a,\rm lin}$ given the observed range of $\Gamma_0\approx 0.05-0.2\muHz$ \citep{Vrard:18}.  The additional uncertainty in $E_{a,\rm lin}$ due to the uncertainty in $\mathcal{P}_0$ is not accounted for in the figure, which assumes the median values for $B$ and $s$ and $\beta=1.5$ (see \citealt{Samadi:12}). The blue region shows the range of possible $E_{\rm th}$ given the order of magnitude range in possible values of $\gamma_a$ (Section~\ref{sec:gamma_a}).  It assumes that there are daughter modes with sufficiently small detuning that $|\Delta_b| \ll \gamma_a$.  The dashed curve shows $E_{\rm th}$ if instead we adopt the likely value of the minimum detuning $|\Delta_b|=\omega_a\ell_b^{-3} n_a^{-2}$, assuming $\gamma_a=10^{-8}\trm { s}^{-1}$ (Section~\ref{sec:Dbc}).

We find that stars with smaller $\nu_{\rm max}$ and larger $M$ are more likely to have an $\ell_a=1$ $p$-$m$ mode with energy  $E_{a,\rm lin}$ above $E_{\rm th}$.  Thus, mixed modes are more likely to be parametrically unstable in  more evolved, more massive stars.  This is because $E_{a,\rm lin}\propto \nu_{\rm max}^{-3.1}$ whereas $E_{\rm th}\propto \nu_{\rm max}^{-2}$ (approximately). Furthermore, at a given $\nu_{\rm max}\ga 50\muHz$, more massive stars have smaller $\gamma_{b,c}$ (i.e., $\alpha$) and larger $\kappa_{abc}$ (see Figure~\ref{fig:gamma_kappa}).  Given the range of plausible values of $E_{a,\rm lin}$ and $E_{\rm th}$, a mixed mode could be unstable out to $\nu_{\rm max} \la 100\muHz$ for $M=1.2 M_\odot$ and $\nu_{\rm max} \la 130\muHz$ for $M=2.0 M_\odot$.  It could be unstable out to even larger $\nu_{\rm max}$ (especially for $M=2.0 M_\odot$) if there are daughters that happen to have especially small detuning of $|\Delta_b| \ll \gamma_a$ or if there are collective sets of unstable daughters (Section~\ref{sec:stability_analytic}). As shown in Figure~\ref{fig:gamma_kappa}, the modes become strongly nonlinear for $\nu_{\rm max}\la 30\muHz$  and the weakly nonlinear stability calculation is no longer applicable (i.e., to the left of the vertical dotted line in Figure~\ref{fig:Eth_Ealin}).

\vspace{1.0cm}
\section{\bf D\lowercase{iscussion}}
\label{sec:discussion}

The amplitudes of mixed modes increase dramatically as a star evolves up the RGB (as $\nu_{\rm max}$ decreases).  The maximum shear of the modes $k_r \bar{\xi}_r$ provides a measure of their nonlinearity. By calibrating to the observed bolometric amplitudes, we showed that the maximum shear $k_r \bar{\xi}_r \propto \nu_{\rm max}^{-9/2}$. Thus, the nonlinearity increases rapidly with decreasing  $\nu_{\rm max}$. We found that the modes are weakly nonlinear ($k_r \bar{\xi}_r \approx 10^{-3}$) by $\nu_{\rm max} \approx 150 \muHz$ and strongly nonlinear  ($k_r \bar{\xi}_r \approx 1$) by $\nu_{\rm max} \approx 30\muHz$, , nearly independent of $M$.  
 
As a mixed mode propagates through the core, its shear increases as $k_r \bar{\xi}_r \propto r^{-2}$, reaching a peak near the inner turning point at  $r \approx 10^{-4} R_\odot (\nu_{\rm max}/100\muHz) $.   A strongly nonlinear wave will break and deposit all of its energy and angular momentum as it approaches the turning point.  By contrast, a weakly nonlinear  wave will, if unstable, excite secondary waves within the core, but only lose a portion of its energy and angular momentum before reflecting at the turning point and propagating back outward.  Although we defer a study of the observational consequences of these effects to future work,  strongly nonlinear waves likely have reduced amplitudes and broadened linewidths.  To a lesser extent, the same might be true of weakly nonlinear waves, although here the calculation is more involved as it depends on the details of the nonlinear saturation by secondary waves.  A full understanding likely requires a large mode network calculation of the type carried out in the context of neutron star $r$-mode instabilities \citep{Arras:03, Brink:05, Bondarescu:09} and  dynamical tides in hot Jupiter systems \citep{Essick:16}.

Interestingly, some red giants exhibit dipole modes with unexpectedly low amplitudes, known as depressed modes  \citep{Mosser:12:depressed, Garcia:14, Stello:16a, Mosser:17}.  Although the prevalence of depressed modes depends on $M$ and $\nu_{\rm max}$, these two parameters alone do not  predict whether a star's dipole modes are depressed. This suggests that an additional stellar property plays a role. \citet{Fuller:15} proposed that some red giants have strong internal magnetic fields that   scatter and trap oscillation-mode energy within the core (the magnetic greenhouse effect). \citet{Stello:16a} find that this mechanism can account for the lack of depressed modes of higher angular degrees (quadrupole and octupole).  However, \citet{Mosser:17} measure the visibilities of depressed modes and find that they are not fully damped in the core, contrary to the predictions of the magnetic greenhouse effect.

Since weakly nonlinear, unstable mixed modes are only partially damped in the core, perhaps the observations by \citet{Mosser:17} indicate that depressed modes are a consequence of weakly nonlinear effects rather than magnetic effects.  Given that mixed modes lie near the parametric instability threshold over a large range of $\nu_{\rm max}$ (see Figure~\ref{fig:Eth_Ealin}), their amplitudes may be  sensitive to details of the individual mode parameters (e.g., mode linewidths, daughter detunings, coupling coefficients) and the complicated, time-dependent nonlinear mode dynamics. This could explain why depressed modes are found to occur over a large range of $\nu_{\rm max}$ and yet two otherwise similar stars (similar $M$ and $\nu_{\rm max}$) might not both exhibit depressed modes.  We found that higher mass stars are more likely to be above the parametric instability threshold for $\nu_{\rm max}\ga 50\muHz$ (compare the top and bottom panels of Figure~\ref{fig:Eth_Ealin}), which is also consistent with observations \citep{Stello:16b}.

Since the nonlinear interactions occur within the core, the degree of amplitude attenuation will depend on the fraction of mode energy that gets transmitted from the acoustic cavity, where the modes are excited, into the $g$-mode cavity (similar to the magnetic greenhouse effect). Even if a mode is damped in the core by nonlinearities, the amplitude attenuation at the surface will be small if the transmitted fraction is small.  This might explain why the visibility of depressed modes increases as $\nu_{\rm max}$ decreases and as the angular degree $\ell$ increases \citep{Mosser:12:depressed, Stello:16a, Mosser:17}.  Mode network calculations are needed in order to assess this explanation.

\vspace{-0.5cm}
\acknowledgements{This work was supported in part by NASA ATP grant NNX14AB40G.  We thank the referee for valuable comments on the manuscript.

\software{\texttt{MESA} \citep[][\url{http://mesa.sourceforge.net}]{Paxton:11, Paxton:13, Paxton:15, Paxton:18},
\texttt{GYRE} \citep[][\url{https://bitbucket.org/rhdtownsend/gyre/wiki/Home}]{Townsend:13, Townsend:18}
}

\vspace{-0.0cm}

\bibliography{refs}

\end{document}